\documentclass[aps,twocolumn]{revtex4}
\usepackage{epsfig}

\begin{document}

\title{Entropy production in the majority-vote model}

\author{Leonardo Crochik and T\^ania Tom\'e}

\affiliation{Instituto de F\'{\i}sica,
Universidade de S\~{a}o Paulo,
Caixa Postal 66318 \\
05315-970 S\~{a}o Paulo, S\~{a}o Paulo, Brazil}

\date{\today}

\begin{abstract}

We analyzed the entropy production in the majority-vote model by
using a mean-field approximation and Monte Carlo simulations.
The dynamical rules of the model do not obey detailed balance 
so that entropy is continuously being produced. 
This nonequilibrium stochastic model is known to have a critical 
behavior belonging to the universality class of the equilibrium Ising model.
We show that the entropy production also
exhibits a singularity at the critical point 
similar to the one occurring in the entropy, or the energy, of the 
equilibrium Ising model.

PACS numbers: 05.70.Ln, 05.50.+q, 02.50.Ga

\end{abstract}

\maketitle

\section{Introduction}

Irreversible systems in the stationary state 
are in a process of continuous production 
of entropy. In systems, like the one  studied here,
governed by a master equation, that is, defined by a 
continuous time Markov process, the irreversibility
is characterized by the lack of detailed balance.
When a system governed by a Markov process is reversible, 
that is, when the dynamic rules are such that detailed
balance is obeyed, the production of entropy vanishes at
the equilibrium state.
This is indeed the case of the Glauber model
and any other dynamics used to simulate the equilibrium Ising model.
The production of entropy is then a signature
of irreversibility. 

The rate of change of the entropy $S$ of a system
can be properly decomposed into two contributions \cite{nico77}
\begin{equation}
\frac{dS}{dt} = \Pi - \Phi,
\label{1}
\end{equation}
where $\Pi$ is the entropy production due to irreversible
processes occurring inside the system and $\Phi$ is the entropy flux
from the system to the environment.
The quantity $\Pi$ is positive definite whereas $\Phi$
can have either sign. In the stationary state 
the entropy $S$ of the system remains constant so
that $\Phi=\Pi$. Notice that the quantity $\Phi$
is defined here as the flux from inside to outside the system.
So that, it will be positive in the stationary state.

In this work we study the steady state production of
entropy in a nonequilibrium lattice model, namely,
the majority-vote model \cite{ligg85,gray85,oliv92}. 
This is a polling
model in which individuals in a community take the opinion
of their neighbors with a certain probability $p$
and the opposite opinion with probability $q=1-p$.
In two or more dimensions, this model displays a continuous 
phase transition described by the same critical exponents
as the equilibrium Ising model \cite{oliv92}.
This is in agreement with the conjecture by 
Grinstein et al. \cite{grin85} 
according to which nonequilibrium stochastic systems with
up-down symmetry fall in the universality class of the
equilibrium Ising model.

The flux of entropy is determined by means of an expression 
which is the average of a function of the rates of
transition from one state to another and its reverse
\cite{maes99,lebo99,maes00,maes03,leco05}.
As the entropy production equals the entropy flux in 
the stationary state, the former can be determined
from the expression for the latter. 
We use mean-field and Monte Carlo simulations to calculate the
entropy flux.
In the stationary regime, and at the critical point, the
entropy flux, or equivalently,the entropy production,
displays a singularity which we assume
to be the same singularity occurring in the entropy of the
equilibrium Ising model.

\section{Entropy production}

Let us consider a system described by a continuous Markov process 
with stochastic variables defined over the sites of a regular lattice. 
A configuration of the system is denoted by 
$\sigma=(\sigma_1,\sigma_2,\ldots,\sigma_N)$ where $N$ is the number
of sites of the lattice and $\sigma_i=\pm1$ is the spin variable
associated to site $i$.
We will be concerned only with one spin flip dynamics,
defined by a transition rate $w_i(\sigma)$ in which
the spin variable $\sigma_i$ changes its sign. 
The time evolution of the probability $P(\sigma,t)$
is governed by the master equation,
\begin{equation}
\frac{d}{dt} P(\sigma,t)=\sum_i 
\{ w_i(\sigma^i) P(\sigma^i,t) -  w_i(\sigma) P(\sigma,t) \},
\label{2}
\end{equation}
where $\sigma^i=(\sigma_1,\sigma_2,\ldots,-\sigma_i,\ldots,\sigma_N)$.

The Gibbs entropy $S(t)$ of the system at time $t$ 
is defined by
\begin{equation}
S(t) = - \sum_\sigma P(\sigma,t) \ln P(\sigma,t).
\label{3}
\end{equation}
Using the master equation (\ref{2}), its 
time derivative can be written as
\[
\frac{d}{dt}S(t) =\frac12 \sum_{\sigma} \sum_i
\ln\frac{P(\sigma^i,t)}{P(\sigma,t)} \times
\]
\begin{equation}
\times \{ w_i(\sigma^i) P(\sigma^i,t) -  w_i(\sigma) P(\sigma,t) \}.
\end{equation}
In agreement with Eq. (\ref{1}),
the righ-hand side of this expression should be decomposed into two terms, 
the entropy  production $\Pi$ and the entropy flux $\Phi$. 
These two quantities have the following expressions 
\cite{maes99,lebo99,maes00,maes03}
\[
\Pi = \frac12\sum_{\sigma} \sum_i
\ln\frac{w_i(\sigma^i)P(\sigma^i,t)}{w_i(\sigma)P(\sigma,t)} \times
\]
\begin{equation}
\times \{ w_i(\sigma^i) P(\sigma^i,t) -  w_i(\sigma) P(\sigma,t) \},
\label{5}
\end{equation}
and 
\begin{equation}
\Phi = \sum_{\sigma} \sum_i w_i(\sigma) P(\sigma,t)
\ln\frac{w_i(\sigma)}{w_i(\sigma^i)}.
\label{6}
\end{equation}
The right-hand side of Eq. (\ref{5}) is always positive as can 
be easily proved,
and the right-hand side of Eq. (\ref{6}) can be written as the average
over the stationary probability distribution, that is,
\begin{equation}
\Phi = \sum_i \langle w_i(\sigma)
\ln\frac{w_i(\sigma)}{w_i(\sigma^i)} \rangle.
\label{7}
\end{equation}

This is a particularly useful equation 
because it can be employed to estimate $\Phi$
from a Monte Carlo simulation. As it
is well known only quantities that can be
written as averages can be determined numerically
in a Monte Carlo simulation. In this sense
it is not possible to determine $S$ given by Eq. (\ref{3})
nor $\Pi$, given by Eq. (\ref{5}),
but it is actually possible to determine $\Phi$
from Eq. (\ref{6}).
We remark finally that in the steady state
$\Pi=\Phi$ so that it is possible to determine
the entropy production in this regime
by a Monte Carlo simulation.

\section{Majority-vote model}

The majority-vote model is a one-spin flip stochastic
dynamics defined by the following transition rate
\begin{equation}
w_i(\sigma) 
= \frac{1}{2} \{ 1-\gamma\sigma_i {\cal F}(\sum_\delta \sigma_{i+\delta}) \},
\label{8}
\end{equation}
where ${\cal F}(x)$ is a function that equals $-1$, $0$ or $+1$
according to whether $x<0$, $x=0$ or $x>0$, and the summation
is over the nearest neighbor sites of site $i$. 
Notice that the transition rate $w_i(\sigma)$ has the
up-down symmetry, that is, it is invariant
under the sign change of all spin variables $\sigma_i$.
At each time interval, a site $i$ is chosen at random.
If the majority of the neighbors are in state $+1(-1)$
then the site takes the value $+1(-1)$ with probability
$p$ and the opposite sign with probability $q=1-p$
where $p=(1+\gamma)/2$.
We will restrict ourselves to the case $0\leq q \leq 1/2$ so that
$1\geq\gamma\geq 0$.
The model can also be interpreted as an Ising system in contact with
two heat reservoir at temperatures $0^+$ and $0^-$.
Putting it in a different way, one reservoir always provides energy 
and the other takes it away. Spin systems in contact with two heat
baths at different temperatures 
\cite{masi85,dick87,garr87,marq89,tome89,tome91,racz94,figu00,leco05}
are perhaps the simplest models with nonequilibrium steady states
exhibiting dynamic phase transitions.

In the stationary regime, the present 
model displays a continuous phase transition from an ordered 
(ferromagnetic) state
to a disordered (paramagnetic) state .
On a square lattice it is found by numerical simulation that
the critical point occurs at $q_c=0.075(1)$ \cite{oliv92}.
The ordered state occurs when $0\leq q <q_c$ and 
the disordered state when $q_c<q\leq 1/2$. 
For $0<q<1/2$, this model does not obey detailed balance
and we expect a strictly positive entropy production.
When $q=1/2$ the system is completely disordered and 
corresponds to a reversible system so that the 
entropy production vanishes in this case.
The critical behavior \cite{oliv92} puts this model in the same
universality class as the equilibrium Ising model.
This result comes from the conjecture by Grinstein et al. 
\cite{grin85} which states that models with stochastic evolution 
rules with up-down symmetry belongs to the Ising universality class.

From the transition rate $w_i(\sigma)$ given by Eq. (\ref{8})
it is straightforward to show that
\begin{equation}
B_i(\sigma) =
\ln\frac{w_i(\sigma)}{w_i(\sigma^i)} =
\left(\ln\frac{q}{p} \right)
\sigma_i {\cal F}(\sum_\delta \sigma_{i+\delta}).
\label{9}
\end{equation} 
Therefore, the entropy flux per site $\phi=\Phi/N$ 
for the majority-vote model can be 
determined as the average
\begin{equation}
\phi = \langle B_i(\sigma) w_i(\sigma) \rangle.
\label{10}
\end{equation}
Notice that for a square lattice the function ${\cal F}(x)$ 
reads
\begin{equation}
{\cal F}(\sigma_1+\sigma_2+\sigma_3+\sigma_4)=
\frac18(\sigma_1+\sigma_2+\sigma_3+\sigma_4)
(3-\sigma_1\sigma_2\sigma_3\sigma_4).
\label{10a}
\end{equation}

\section{Mean-field results}

%------------------ Figure 1 --------------------
\begin{figure}[t]
\epsfig{file=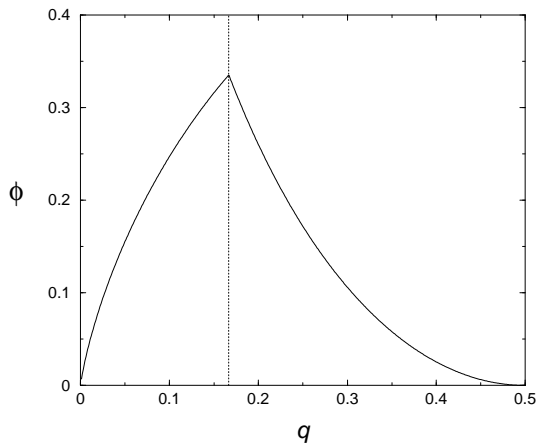,height=6cm}
\caption{Entropy flux $\phi$ for the majority-vote model
in the simple mean-field approximation as a function
of the external parameter $q$. 
The dotted line indicates the position of the 
critical point occurring at $q_c=1/6$.}
\end{figure}
%------------------------------------------------

From the master equation we get the following equations
for the time evolution of the magnetization $\langle\sigma_i\rangle$
\begin{equation}
\frac{d}{dt} \langle\sigma_i\rangle =
-2\langle\sigma_i w_i(\sigma)\rangle.
\end{equation}

In the first order dynamic mean-field approximation,
or simple mean-field approximation,
the correlations are neglected and we need only this equation. 
We apply the approximation to the case of a regular
lattice of coordination four.
In the stationary state, the magnetization 
$m=\langle\sigma_i\rangle$ is given by the equation
\begin{equation}
m= \frac{\gamma}{2} m (3-m^2).
\end{equation}
Using this approximation, we derive the following expression for
the entropy flux
\begin{equation}
\phi = \left(\ln\frac{q}{p}\right)
\{ \frac14(3m^2-m^4)-\frac{\gamma}{16}(5+6m^2-3m^4) \}.
\end{equation}

The  paramagnetic solution, $m=0$, 
gives the following expression for the
entropy flux in the paramagnetic phase
\begin{equation}
\phi = \frac{5}{16}(1-2q)\ln\frac{1-q}{q}.
\end{equation}
The ferromagnetic solution is given by the expression
\begin{equation}
m = \sqrt{\frac{1-6q}{1-2q}},
\end{equation}
which is valid for $q<q_c=1/6$.
From this result it follows that the entropy flux 
in the ferromagnetic state is
\begin{equation}
\phi = \frac{q(1-q)}{1-2q}\ln\frac{1-q}{q}.
\end{equation}
The stationary entropy flux, or equivalently the entropy production,
is a continuous function of the
parameter $q$ as shown in Fig. 1. At the critical 
point it presents a singularity represented
in this mean-field approximation by a discontinuity in the first
derivative. 

\section{Numerical simulations}

We have simulated the majority-vote model on a square
lattice with periodic boundary conditions for
different values of the size $N=L\times L$ of the system. 
The simulation was performed as follows. At each time
step a site is chosen at random. It takes the
value of the majority sign of its neighbors with  probability
$p=1-q$ and the opposite sign with probability $q$
in accordance with the prescription given by Eq. (\ref{8}).
After discarding the first Monte Carlo steps
the stationary properties are calculated.
We used from $10^6$ to $10^7$ Monte Carlo steps
to calculate the averages such as the flux $\phi$
given by Eq. (\ref{10}). The magnetization and other 
quantities such as the susceptibility have
already been determined by Monte Carlo simulations \cite{oliv92}
and will not concern us here. It is found that
a continuous phase transition takes place at
$q_c=0.075(1)$ and that the critical exponents
are the same as those of the two-dimensional Ising model \cite{oliv92}.

%------------------ Figure 2 --------------------
\begin{figure}[t]
\epsfig{file=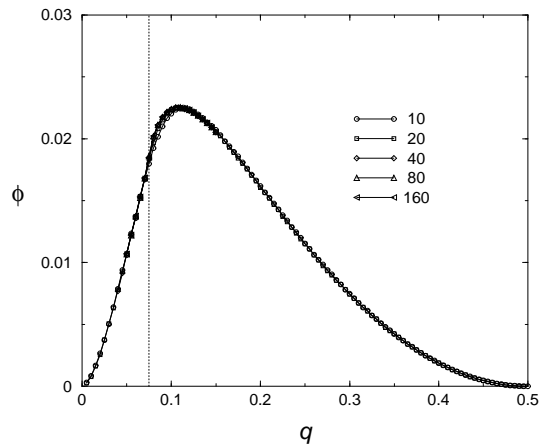,height=6cm}
\caption{Stationary entropy flux $\phi$ for the majority 
vote model by Monte Carlo simulation as a function
of the external parameter $q$ for various values
of the system size $L$. 
The dotted line indicates the position of the 
critical point occurring at $q_c=0.075$.}
\end{figure}
%------------------------------------------------

%------------------ Figure 3 --------------------
\begin{figure}[t]
\epsfig{file=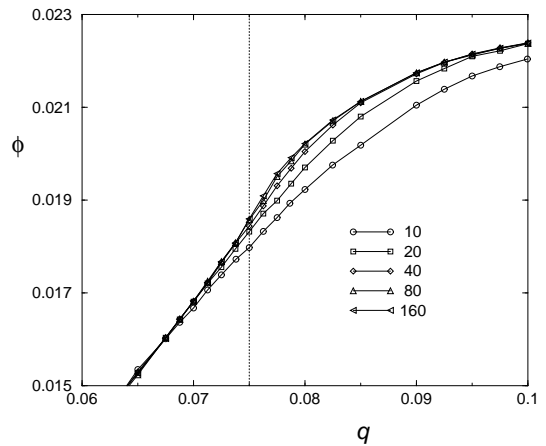,height=6cm}
\caption{Enlargement of Fig. 2 around the critical point.}
\end{figure}
%------------------------------------------------

In Fig. 2 we show the numerical results for
the entropy flux $\phi$ for several values
of the size of the system $L$.
The entropy flux is finite and continuous.  
It has a maximum and vanishes when $q\to 1/2$ and when
$q\to 0$ as expected since in these two limits
the system reaches an equilibrium stationary state.
When $q\to 1/2$ we found numerically
that the flux vanishes according to 
\begin{equation}
\phi = b (\frac12-q)^2,
\end{equation}
with $b=0.190(3)$ and when $q\to 0$, it vanishes
according to 
\begin{equation}
\phi = a q^2 \ln\frac{1-q}{q},
\end{equation}
with $a=1.83(5)$.

We remark that  the critical point
does not correspond to the maximum of $\phi$. 
Actually, it corresponds to the point of inflexion
occurring just before the maximum as can be seen in Fig. 3. 
At the critical point the flux is finite but
has a singularity  which we assume to be 
the same type as that of the energy or the entropy of 
the equilibrium Ising model, namely, 
of the form
\begin{equation}
\phi = \phi_c + A_{\pm}|q-q_c|^{1-\alpha},
\label{19}
\end{equation}
where $\phi_c$ is the value of the entropy flux
at the critical point.
The amplitudes $A_+$ and $A_-$ correspond
to the regimes below ($q<q_c$) and above ($q>q_c$)
the critical point. 

The energy of the equilibrium Ising model
is related to the short range correlations of the even type.
From Eqs. (\ref{9}) and (\ref{10}) and the result given by Eq. (\ref{10a})
we see that the entropy flux is also related
to the short range correlations of the even type: 
$\langle\sigma_i\sigma_j\rangle$
and $\langle\sigma_i\sigma_j\sigma_k\sigma_\ell\rangle$
where $i,j,k$ and $\ell$ are nearest and next-nearest neighbor
sites. We expect, therefore, that the critical behavior of the entropy
flux of nonequilibrium models with up-down symmetry
and the critical behavior of energy of the equilibrium
Ising model are described by the same critical exponent.

%------------------ Figure 4 --------------------
\begin{figure}[t]
\epsfig{file=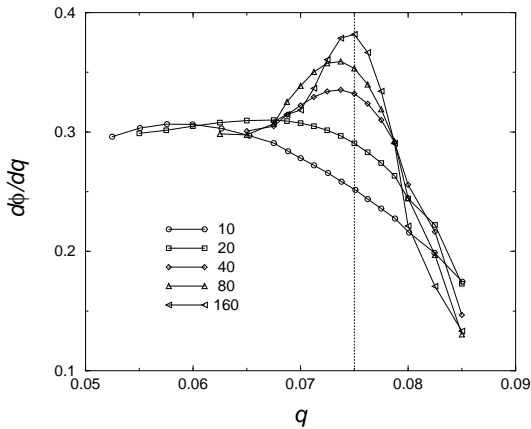,height=5.8cm}
\caption{Derivative $d\phi/dq$ of the entropy flux
$\phi$ with respect to the external parameter $q$
for several values of the system size $L$.
The dotted line indicates the position of the 
critical point occurring at $q_c=0.075$.}
\end{figure}
%------------------------------------------------

%------------------ Figure 5 --------------------
\begin{figure}[t]
\epsfig{file=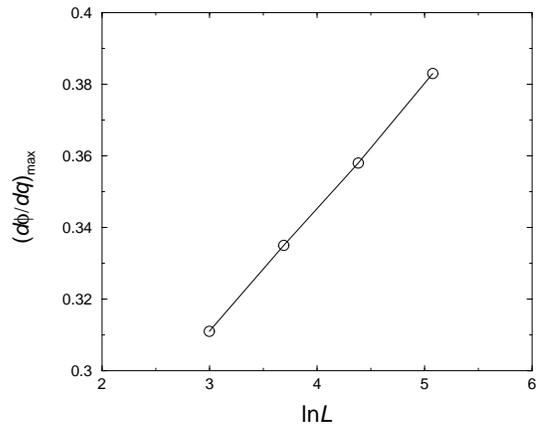,height=5.8cm}
\caption{Maximum of the derivative $d\phi/dq$ of the entropy flux
$\phi$ with respect to the external parameter $q$
as a function of the logarithm of the system size $L$.
}
\end{figure}
%------------------------------------------------

To determine the critical behavior it is convenient 
to study the derivative of the entropy flux
with respect to the external parameter $q$
which, as follows from  Eq. (\ref{19}), behaves as
\begin{equation}
\frac{d\phi}{dq} \sim |q-q_c|^{-\alpha}.
\end{equation}
The exponent $\alpha$ is the same exponent
related to the specific heat of the equilibrium
Ising model. Since, for the square lattice,
the singularity is of the logarithm ($\alpha=0$) type
then we assume that for the present two-dimensional
majority-vote model
\begin{equation}
\frac{d\phi}{dq} \sim |\ln|q-q_c||.
\label{21}
\end{equation}

To test the assumption given by Eq. (\ref{21})
we have numerically determined $d\phi/dq$ for several lattice
sizes $L$ as shown in Fig. 4.
Using a finite-size scaling theory \cite{ferd69}, then
this quantity as a function of $L$ should diverges as
$\ln L$
%\begin{equation}
%\left(\frac{d\phi}{dq}\right)_c \sim \ln L
%\end{equation}
at the critical point and a similar behavior
at the maximum, that is,
\begin{equation}
\left(\frac{d\phi}{dq}\right)_{\rm max} \sim \ln L.
\end{equation}
From Fig. 5 we see that this behavior is indeed
followed when $L\ge20$.

\section{conclusion}

We have determined by mean-field approximation and by 
Monte Carlo simulations the stationary entropy production 
of a nonequilibrium
model with up-down symmetry, namely, the majority-vote model.
The calculation of this quantity by means of Monte Carlo simulations
was possible because it equals the entropy flux, in the 
stationary state, and so this quantity can
be written as an average over a
stationary probability distribution.
The mean-field analysis as well as the Monte Carlo simulations
show that the stationary entropy production
is positive for nonequilibrium situation, it
vanishes when the system attains an equilibrium
stationary state, and
exhibits a singularity at the critical point.
The mean-field results gives a
singularity represented by a discontinuous first derivative
as usually happens in mean-field calculations of the energy
as a function of temperature in equilibrium spin models.
The Monte Carlo data, analyzed by a finite-size scaling theory,
have shown that the stationary entropy production has the same singular
behavior at the critical point as the energy of the
equilibrium Ising model. In the present case, namely,
the model defined on a square lattice, the singularity
of the derivative of the entropy production
is characterized by a logarithmic divergence.

\section*{Acknowledgment}

T. T. and L. C. acknowledge financial support from CNPq 
and FAPESP, respectively.

\end{document}